\documentstyle[nanosymp,epsf]{article}
\begin{document}

\title{Pseudo-gap and spin polarization in a two-dimensional \\electron gas}

\author{{\em V.~V'yurkov}\dag, A.~Vetrov\ddag, and V.~Ryzhii}
\affil{University of Aizu, Aizu-Wakamatsu 965-8580, Japan\\
\ddag Institute of Physics and Technology, RAS, Nakhimovsky prosp. 34, Moscow, 
117218, Russia}
\footnote{\dag On leave from the Institute of Physics and Technology, RAS, 
Moscow, Russia}

\beginabstract

Tunnelling density of states in the vicinity of Fermi level of a 
two-dimensional electron gas subjected to an external 
parallel and zeroth magnetic field is calculated. It reveals a pseudo-gap 
recently 
observed in the experiments. The gap originates in spin polarization of 2DEG. 
Non-monotonic dependence of energy on a Landau level filling factor (density) 
was obtained. It implies the tunneling current peculiarities at filling factors 
1/2 and 1. The Ising-like model 
of the exchange interaction in 2DEG was exploited instead of the conventional 
one. It was crucial to achieve even a qualitative agreement with experimental 
data.

\endabstract

\section*{Introduction}

The investigation of a two-dimensional (2DEG) is still in focus of modern 
physics. Lately spin effects attracted a great attention of researchers. 
Spontaneous spin polarization, partial as well as total one, along with a 
polarization induced by external magnetic 
field are widely discussed. 
Some peculiarities of tunneling current were observed recently. Evidently, they 
are closely connected with unusual 
behavior of density of states (DOS) by the Fermi level in 2DEG.
There are two groups of experiments, the first one deals with parallel and zero 
magnetic field [1] and the second one investigates perpendicular magnetic field, 
that is, the regime of the Quantum Hall Effect [2-4].
 
Lately the tunneling spectroscopy revealed two peaks in DOS surrounding the 
lowered DOS at the Fermi level [4]. 
The authors treated this behavior as a manifestation of the Coulomb pseudo-gap 
predicted by 
Efros and Shklovskii long time ago. 
The crucial feature making this explanation doubtful is a dependence of the 
pseudo-gap shape upon 
external magnetic field. In the contrary, in the experiment [3] the observed 
peculiarities of DOS were 
regarded as an indication to spin polarization.   
Here the effect of spin polarization on the tunneling current is discussed.
A similar approach was previously used to obtain 
a lowered DOS at the Fermi level in 1DEG [5] and 2DEG [6]. In particular, it 
results in a lower conductance step with 
respect to a conductance quantum in a quantum wire [5].   

\section {Exchange interaction model for zero and parallel magnetic field}
Throughout the paper we employ the Ising model to describe the exchange 
interaction and therefore use the following relation to calculate the exchange 
energy 
\begin{equation}
E_{ex}  = - {e \over 2}\sum\limits_{kq \ne 0\sigma } {V(q) } n_{k\sigma } n_{k + 
q\sigma }  
+ e\sum\limits_{kq \ne 0} {V(q)} n_{k\sigma  =  + } n_{k + q\sigma  =  - }
\end{equation} 
,where
$
n_{k\sigma } 
$
=0,1 is a filling factor of the electron state with the wave vector k and spin 
orientation 
$
\sigma= + (up), - (down)
$
, $n$ is an electron sheet density, $V(q)$ is a 2D Fourier transform of the 
Coulomb potential. For the unscreened Coulomb potential $V(q)=2\pi e/ \kappa q$, 
where $\kappa$ is a permittivity. 

\begin{figure}[b]

\epsfbox{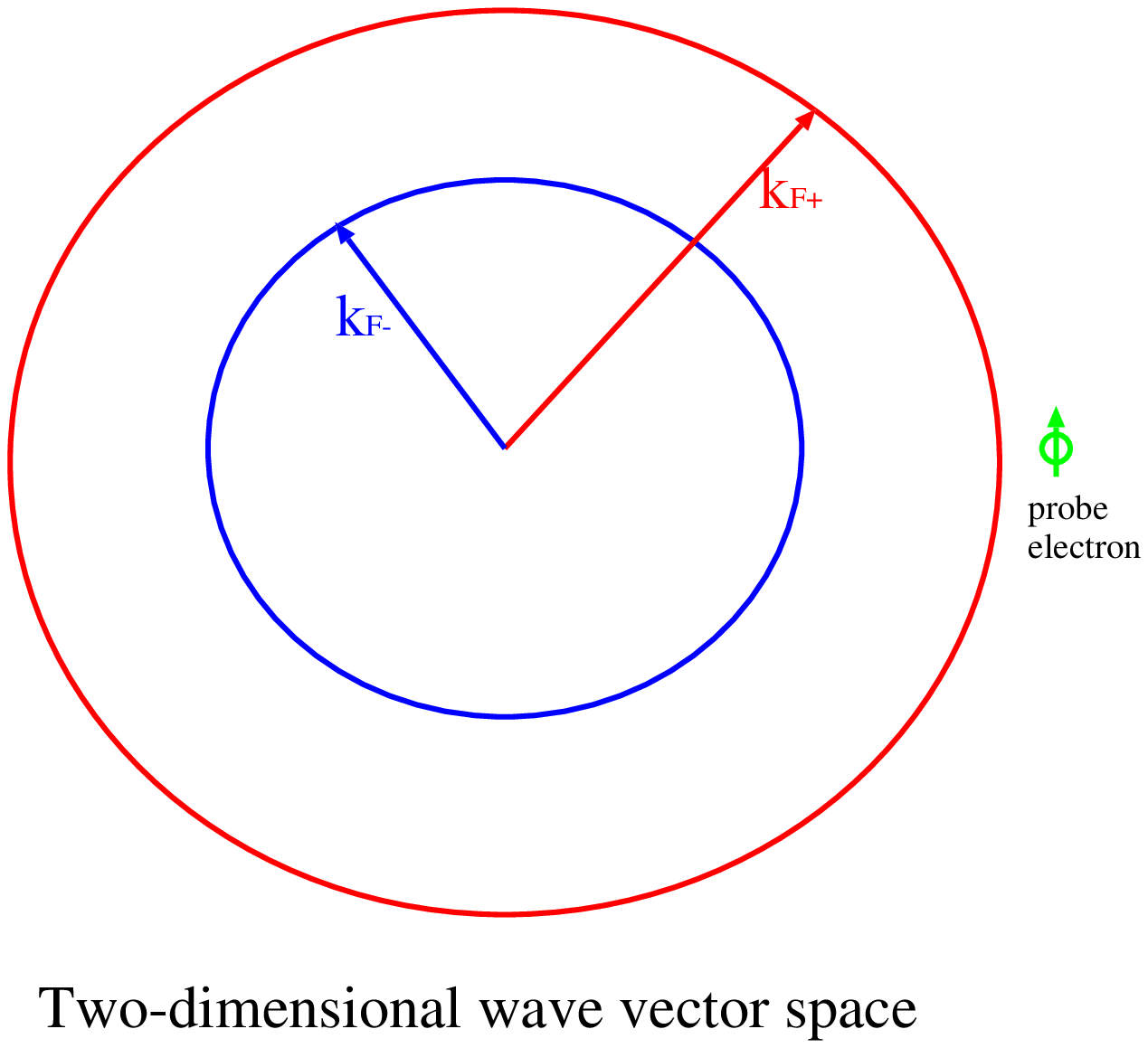}
\caption{The 2D wave vector space with the Fermi wave vectors $k_{F+}$ and 
$k_{F-}$ for majority (+) and minority (-) spin polarization, respectively.}
\end{figure}
The second term regarding the interaction between electrons with opposite spin 
orientation differences the Ising (I) model from that of Kohn-Sham (KS). The I-
model turned out to be more adequate to the available experimental data. 

In Fig.1 the spin polarized state of a 2DEG is depicted.

The total energy should acquire the Zeeman term if the 2DEG is embedded in the 
external magnetic field B. 
\begin{equation}
E_B  =  - {1 \over 2}g_0 \mu _B B(\sum\limits_{k\sigma  =  + } {n_{k\sigma }  - 
} \sum\limits_{k\sigma  =  - } {n_{k\sigma } )} 
\end{equation}
The ground state of the system comprising filling factors 
$
n_{k\sigma }  = 0
$
 or 1 for the fixed electron sheet density $n= n_{+} + n_{-}$  is obtained on 
minimizing the total energy including kinetic, direct Coulomb, exchange, and 
magnetic energy.
The correlation energy was ignored as the leading term was supposed to be that 
of exchange. 
At least, an exchange interaction was sufficient to qualitatively explain a 
pseudo-gap shape and its dependence on a 
magnetic field.
\section {Tunnelling density of states}
Here we only refer to the tunnelling density of states (TDOS) 
$
D(\varepsilon )
$
 which could be obtained in tunnel spectroscopy experiments.
 The tunnel current between two electron reservoirs biased with a voltage 
$
V
$
is supplied by a convolution (see, for instance, [4])
\begin{equation}
I \sim \int\limits_{ - \infty }^\infty  {D_l (\varepsilon )D_r (\varepsilon  - 
eV)\left[ {f_r (\varepsilon  - eV,T) - f_l (\varepsilon ,T)} \right]d\varepsilon 
} 
\end{equation} 
Here 
$
f
$
 is the Fermi distribution function for temperature 
$
T
$
, 
the subscript 
$
r
$
 corresponds to the right reservoir while the subscript l corresponds to the 
left one.

For spin-dependent tunneling one must employ the above expression separately for 
spin-up and spin-down electrons. 
When one reservoir is a metal or heavily doped semiconductor its 
DOS may be replaced by a constant. Then the DOS associated with the reservoir of 
2DEG at zero temperature 
$
T = 0
$
 can be obtained from a derivative 
$
{\rm{dI/dV}}
$
. 

For weak tunneling events the electrons added to the system or removed from it 
do not much disturb the state of the system. 
The tunnel DOS could thus be calculated in the following way.
 Firstly, one should calculate the energy 
$
\varepsilon {\rm{(k,}}\sigma {\rm{) }}
$
required to add an electron into empty state 
$
{\rm{(k,}}\sigma {\rm{) }}
$
above the Fermi level or remove an electron from the occupied state 
$
{\rm{(k,}}\sigma {\rm{) }}
$
under the Fermi level. It corresponds to different bias polarity. The derivative 
$
{\rm{d}}\varepsilon {\rm{/dk }}
$
gives the TDOS of 2DEG
\begin{equation}
{\rm{D(}}\varepsilon {\rm{) \sim  k |d}}\varepsilon {\rm{/dk|}}^{{\rm{ - 1}}} .
\end{equation} 
Worth mentioning the derivative 
$
{\rm{d}}\varepsilon {\rm{/dk }}
$
may be large although the corrections to the energy caused by exchange 
interaction are small. 
One more remark is that calculations of $D(\varepsilon)$
should be done rather carefully as the reciprocal function $k(\varepsilon)$ 
could be manifold.  

The exchange energy of a probe electron with a spin $\sigma$
and a wave vector 
$
{\rm{k}}
$
 is calculated according to the relation 
\begin{equation}
\varepsilon (k,\sigma  = \pm ) = \pm \sum\limits_{q \ne 0} {V(q)n_{k + q,\sigma  
=  + } }  \mp \sum\limits_{q \ne 0} {V(q)n_{k + q,\sigma  =  - } }
\end{equation} 
,where filling factors 
$
n_{k,\sigma } 
$
 correspond to the ground state configuration. 
It is readily seen from above equation that the interaction with electrons of 
the major
 spin orientation in the interval 
$
(k_{F+} ,k_{F-} )
$
in 2D k-space determines the exchange energy. The exchange energy is thus 
roughly proportional to a polarization degree
$
|k_{F+}  - k_{F-} |
$
which depends on a magnetic field.

\begin{figure}[t]
\leavevmode
\epsfbox{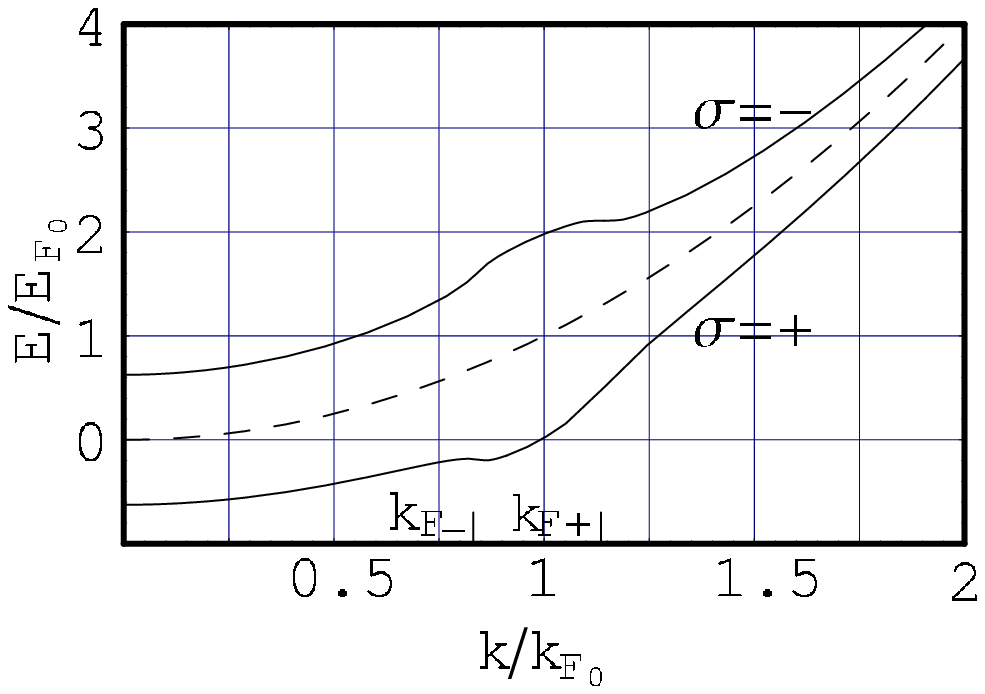}
\caption{The calculated dependence of the probe electron energy on its wave 
vector.}
\end{figure}

In Fig.2 the calculated dependence of the probe electron energy on its wave 
vector for electron density $n = 1.5 \times 10^{11} cm^{ - 2} $ and a magnetic field 
$B=2T$ for a 2DEG in GaAs/AlGaAs heterostructure is plotted. It exhibits the 
derivatives $d\epsilon/dk$=0 around both $k_{F+}$ and $k_{F-}$. This 
corresponds to the infinite density of states. Evidently, screening of Coulomb 
potential smoothens these singularities. However, these peculiarities explain 
the associated peculiarities in tunneling current observed in [1].     

\section {Perpendicular magnetic field}

We evaluated the dependence of energy $\epsilon$ of a 2DEG subject to a 
perpendicular magnetic field on the Landau filling factor $\nu$ via the 
relations
\begin{equation}
\epsilon(\nu)= \frac{\hbar \omega_{c}}{2} \nu - 
A \nu^{3/2} \frac{e^{2}}{\kappa L} exp (-1/2\pi \nu),
\end{equation}
for $\nu  \le {1 \over 2}$ (when electrons with the same spin polarization occupy the 
first Landau level) and
\\
$$\epsilon(\nu)= \epsilon(\nu=1/2)+(1/2)\nu\frac{\hbar\omega_{c}}{2}(\nu-1/2)
+(A/2) \nu^{1/2} \frac{e^{2}}{\kappa L} exp (-\pi)$$
\begin{equation}
- A (\nu-1/2)^{3/2} \frac{e^{2}}{\kappa L} exp (-1/2\pi (\nu-1/2))
\end{equation}
,for ${1 \over 2} \le \nu  \le 1$ (when electrons with the opposite spin polarization begin 
to occupy the first Landau level), $A$ is a numerical constant of the order 
order of unity.  
Here $\hbar$ and $L$ are the cyclotron frequency and the quantum Larmore radius, 
respectively. In general, the Zeeman term analogous to that in Eq.(2) should be 
included. But it turned out to be small compared to the kinetic and exchange 
energies. 
The Fig.3 demonstrates how the energy $\epsilon$ depends on a Landau level 
filling factor $\nu$ (i.e., density). Providing the tunneling current is 
proportional to the derivative $d n/ d\epsilon$ the non-monotonic behavior of 
this dependence implies the existence of singularities of tunneling current for 
$\nu$=1/2 and $\nu$=1. This effect at some rate reminds the Fractional Quantum 
Hall Effect. The occurrence of non-monotonic function $\epsilon (\nu)$ merely 
originates in strong exponential dependence of the exchange overlap integral on 
the distance between interacting electrons. 

\begin{figure} 
[t]\epsfbox{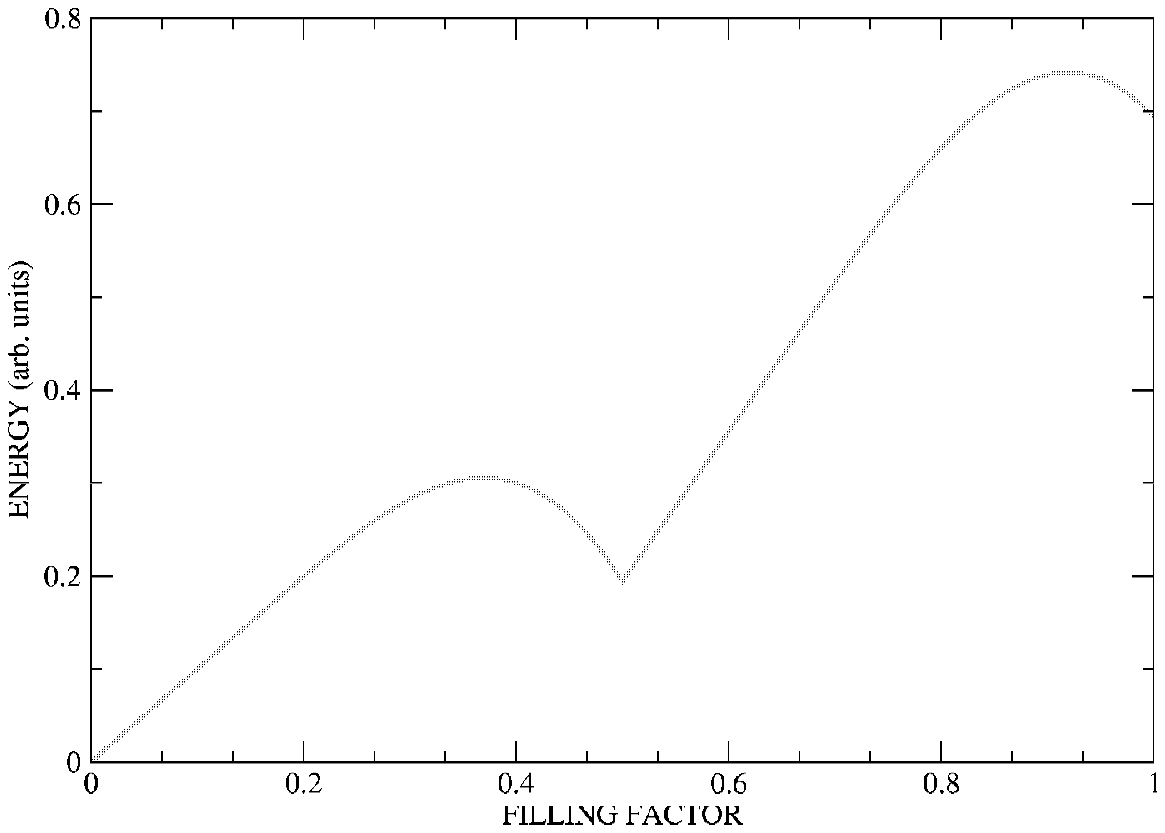}

\caption{The dependence of 2DEG energy $\epsilon$ on the filling factor $\nu$ 
for B=10T.}
\end{figure}

\section{Conclusion} 

Tunnelling density of states in the vicinity of Fermi level of a 
two-dimensional electron gas subjected to an external 
parallel and zero magnetic field is calculated. It reveals a pseudo-gap recently 
observed in the experiments. The gap originates in spin polarization of 2DEG. 
Non-monotonic dependence of energy on a Landau level filling factor (density) 
was also obtained. It makes possible the tunneling current peculiarities at 
filling factors 1/2 and 1. The Ising-like model 
of the exchange interaction in 2DEG exploited instead of the conventional one is 
crucial to achieve even a qualitative agreement with experimental data.

\end{document}